\documentclass[a4paper,10pt,twocolumn]{elsarticle}

\usepackage{mathrsfs,amsmath,amssymb,multicol}
\usepackage[numbers]{natbib}
\biboptions{numbers,sort&compress}
\journal{Prepring}

\begin{document}

\title{ Multi-component lattice Boltzmann models for accurate simulation of flows with wide viscosity variation}

\author{Hiroshi Otomo}\ead{hotomo@exa.com}\author{Bernd Crouse, Marco Dressler, David M. Freed, Ilya Staroselsky, \\ Raoyang Zhang, Hudong Chen }

\address{Exa Corporation, 55 Network Drive, Burlington, Massachusetts 01803, USA}
\cortext[mycorrespondingauthor]{Corresponding author at Exa Corporation}

%}


%\author{Author1}\ead{hotomo@exa.com}\author{Author2}
%\address{Exa Corporation, 55 Network Drive, Burlington, Massachusetts 01803, USA}

%\date{\today}

\twocolumn[
  \begin{@twocolumnfalse}

\begin{abstract}
Multi-component lattice Boltzmann models operating in a wide range of fluid viscosity values are developed and examined.
The algorithm is constructed with the goal to enable engineering applications without sacrificing simplicity and computational efficiency present in the original Shan-Chen model and D3Q19 lattice scheme.
Boundary conditions for modeling friction and wettability effects are developed for discrete representation of surfaces within a volumetric approach, which results in accurate flow simulation in complex geometry. 
Numerical validation of our models includes comparison to previous studies and analytical solutions. The results are shown to be robust and accurate up to an extremely small kinematic viscosity value of $0.0017$ lattice units and the extremely high ratio of components' kinematic viscosities of hundreds and up to a thousand. This improvement is significant compared to previous studies with Shan-Chen model \cite{2013_Yang,2004_Kang,2010_Dong}, in which reasonable accuracy was kept only at the viscosity ratio up to 10 in the Poiseuille flow and the fingering simulation.
\end{abstract}
\maketitle
\end{@twocolumnfalse}
  ]

% \pacs{Valid PACS appear here}% PACS, the Physics and Astronomy
% \keywords{Suggested keywords}%Use showkeys class option if keyword

%%%%%%%%%%%%%%%%%%%%%%%%%%%%%%%%%%%%%%%%%%%%%%%%%%%%%%%%%%%%%%%%%%%%%%%%%%%%%%%%%%%%%%%%%%%%%%%%%%%%%%%%%%%%%%%%%%%%%%%%%%%%%
%
%Introduction
%
%%%%%%%%%%%%%%%%%%%%%%%%%%%%%%%%%%%%%%%%%%%%%%%%%%%%%%%%%%%%%%%%%%%%%%%%%%%%%%%%%%%%%%%%%%%%%%%%%%%%%%%%%%%%%%%%%%%%%%%%%%%%%

%%%%%%%%%%%%%%%%%%%%
\section{Introduction}
%%%%%%%%%%%%%%%%%%%%

The dynamics of multi-component flow commands considerable attention in many fields of physics and engineering due to its fundamental importance and industrial applications. In the recent decades, computational simulation has become a powerful tool to study multi-component flow, with the lattice Boltzmann method gaining popularity among numerous numerical techniques because it combines the microscopic and  macroscopic approaches, its adaptability for complex geometry, and computational efficiency.

For the engineering applications, the adequate account of fluid properties such as viscosity and surface tension is required. If those can be varied at multiple resolutions and simulated Mach numbers without sacrificing accuracy and stability, one can accurately simulate a variety of systems within the appropriate range of characteristic non-dimensional parameters such as the Bond, Capillary, and Weber numbers, which is usually important for the multi-component flow.
In particular, maintaining accuracy and stability of the simulation for a wide range of fluid viscosity variation is a very important functionality required for simulation of many flows, such as flow through the rock where numerical matching of the Capillary number may be necessary for accurate prediction of industrially important parameters.

One of the most popular multi-component LB models is known as the Shan-Chen (SC) model or, including its generalizations, the pseudo-potential model. While widely used, this model operates in a very limited range of viscosity values. 
For an example, in Ref.~\cite{2013_Yang} it is reported that the original SC model is only stable within the viscosity ratio between components varying up to 5. 
While a recent enhancement described in \cite{2012_Porter} improves the stable viscosity ratio all the way into the range up to 1000, the model of Ref.~\cite{2012_Porter} is too complicated to deliver computational efficiency and universality. In particular, its enhanced isotropy requires increased stencils which causes high computational cost and complicates application in arbitrarily complex geometry, while the multiple relaxation time scheme hurts implementation and performance.
In the present study, the further development of SC model is driven by requirements of engineering applications. In particular, retaining the single-relaxation time scheme and the fourth isotropy order which can be naturally achieved by D3Q19, we modify the collisional operator in order to better control instabilities. We also introduce boundary conditions using a volumetric-based approach for modeling the friction and wettability effects.

This paper is organized as follows.
In Section.~\ref{sec:LB_formalism}, our modeling approach is discussed and some details of our LB models are presented. The comprehensive testing aimed at validating the model for engineering application requirements is described in Section.~\ref{sec:validation}. These tests include a static droplet in free space, a multi-component Poiseuille flow, a slug between flat plates, a droplet on an inclined wall, $1 \%$ mixture flow in packed spheres, displacement of a slug from a channel, and fingering. In Section.~\ref{sec:summary}, we summarize our observations and conclusions. In all cases of this study, density ratio between componets is set as 1. 

%%%%%%%%%%%%%%%%%%%%
\section{Lattice Boltzmann models for immiscible fluids}
\label{sec:LB_formalism}
%%%%%%%%%%%%%%%%%%%%

An inter-particle lattice Boltzmann model for immiscible fluids with wide viscosity range is introduced. The model is based on the Shan-Chen model \cite{1993_Xiaowen,1994_Xiaowen} and its recent advancements \cite{2006_Chen, 2006_Zhang, 2006_Latt, 2006_Xiaowen_JFM, 2012_Qli}. The algorithm for bulk solver is presented first and followed by boundary conditions including surface friction and wetting condition on arbitrary geometry.

%Combined with other recent LBM advancements \cite{2006_Chen, 2006_Zhang, 2006_Latt, 2006_Xiaowen_JFM, 2012_Qli}, our model provides accurate and stable results, in particular for small viscosity and in arbitrary geometry. The formalism that we use is briefly described below.

The lattice Boltzmann (LB) equation for multi-component fluid is: 
\begin{equation}
\label{eq:basic_LBM_eq}
 f_{i}^{\alpha} \left( \vec{x}+\vec{c}_{i} \Delta t, t+\Delta t \right) - f_{i}^{\alpha} \left( \vec{x} , t \right)
= \mathcal{C}_{i}^{\: \alpha} + \mathcal{F}_{i}^{\alpha},
\end{equation}
where $f_i^{\alpha}$ is the density distribution function of fluid component $\alpha$ and $\vec{c}_{i}$ is the discrete particle velocity. Henceforth, for simplicity, binary fluid is considered, namely $\alpha=\left\{1, 2 \right\}$, although the framework of this study can be extended to arbitrary number of components. The D3Q19 \cite{1992_Qian} lattice model with the fourth order lattice isotropy is chosen in this study.
The term $\mathcal{F}_{i}^{\alpha}$ is associated with inter-component interaction force and its details are discussed in the follows.
The collision operator $\mathcal{C}_{i}^{\: \alpha}$ is the Bhatnagar-Gross-Krook type,
\begin{equation}
\label{col_operator}
\mathcal{C}_{i}^{\: \alpha} = -\frac{1}{\tau_{mix}}(f_{i}^{\alpha} - f_{i}^{eq, \alpha}).
\end{equation}
Here, $f_{i}^{eq, \alpha}$ is the equilibrium distribution function for the Stokes flow with the third order expansion in $\vec{u}$ ,
\begin{equation}
\label{eq:feq}
f_{i}^{eq, \alpha} = \rho_{\alpha} w_{i} \left[ 1 + \frac{\vec{c}_{i} \cdot \vec{u}}{T_0}  +  \frac{\left( \vec{c}_{i} \cdot \vec{u} \right)^3}{6T_0^3} -  \frac{\vec{c}_{i} \cdot \vec{u}}{2T_0^2}\vec{u}^2 \right],
%f_{i}^{eq, \alpha}(\rho_{\alpha}, \vec{u}) = \rho_{\alpha} w_{i} \left[ 1 + \frac{\vec{c}_{i} \cdot \vec{u}}{T_0}  + \frac{\left( \vec{c}_{i} \cdot \vec{u} \right)^2}{2T_0^2} - \frac{\vec{u}^2}{2T_0} +  \frac{\left( \vec{c}_{i} \cdot \vec{u} \right)^3}{6T_0^3} -  \frac{\vec{c}_{i} \cdot \vec{u}}{2T_0^2}\vec{u}^2 \right],
\end{equation}
where $T_0 = 1/3$ and $w_i$ denote the lattice temperature and isotropic weights in D3Q19, respectively. 
The density of the component $\alpha$, $\rho_{\alpha}$, and the mixture flow velocity, $\vec{u}$, are defined as,
\begin{eqnarray} 
\rho_{\alpha} = \sum_i f_i^{\alpha}, \; \; \;
\rho = \sum_{\alpha} \rho_{\alpha} = \sum_{\alpha} \sum_i f_i^{\alpha}, \; \; \; \nonumber \\
\vec{u^{\alpha}} = \sum_i \vec{c}_{i} \cdot f_i^{\alpha} / \rho_{\alpha}, \; \; \;
\vec{u} = \sum_{\alpha} \sum_i \vec{c}_{i} \cdot f_i^{\alpha} / \rho.
\end{eqnarray}
The relaxation time $\tau_{mix}$ in Eq.~(\ref{col_operator}) relates to the kinematic viscosity of the mixture of components, $\nu_{mix}$, as
\begin{equation}
\tau_{mix} = \left( \nu_{mix} / T_0  \right) + 1/2,
\end{equation}
\begin{equation}
\label{mix_nu}
\nu_{mix}  = p \nu_{1}+ (1- p ) \nu_{2}, 
% \left( \rho_{1} \nu_{1}+\rho_{2}\nu_{2} \right)/ \left(\rho_{1}+\rho_{2}\right).
\end{equation}
where $p$ is the smooth function of $\rho_1$ and $\rho_2$, which ranges from 0 to 1. 
Following the conventional way \cite{1993_Xiaowen,1994_Xiaowen}, the inter-component force, $\vec{F}^{\alpha, \beta}$, between component $\alpha$ and $\beta$ is defined as, 
\begin{equation}
\label{eq:comp_interaction}
\vec{F}^{\alpha, \beta} \left( \vec{x} \right) = G \rho_{\alpha} \left( \vec{x} \right) \sum_{i} w_{i} \vec{c}_{i} \rho_{\beta} \left(  \vec{x}+ \vec{c}_{i} \Delta t \right),
\end{equation} 
for $\alpha \ne \beta$ and  $\vec{F}^{\alpha, \beta} \left( \vec{x} \right) =0$ for $\alpha=\beta$. When the interaction strength $G$ is negative, repulsive force acts between components and yields the separation.
Following the manner in \cite{2012_Qli}, this inter-component force is implemented to the forcing term $\mathcal{F}_{i}^{\alpha}$ in Eq.~(\ref{eq:basic_LBM_eq}).
The acceleration of the component $\alpha$, $\vec{g}_{\alpha}$, originated from $\vec{F}^{\alpha, \beta}$ is defined by $\vec{g}_{\alpha}= \sum_{\beta} \vec{F}^{\alpha, \beta} / \rho_{\alpha}$. The resulting fluid velocity $\vec{u}_{F}$ is defined as the velocity averaged over pre- and post- collision steps and written as,
\begin{eqnarray}
\vec{u}_{F}= \vec{u}+\vec{g}\Delta t /2,  \; \; \; \; \;
\vec{g} = {\sum_{\alpha} \vec{g}_{\alpha}\rho_{\alpha}} / \rho.
\end{eqnarray}
In what follows, this quantity $\vec{u}_{F}$ is called simply \emph{velocity}.

In order to enhance stability and accuracy when $\tau_{mix}$ is not close to 1, effects from the non-equilibrium state are regulated.
After rearrangement of Eq.~(\ref{eq:basic_LBM_eq}), one obtains,
\begin{equation}
\label{eq:LBM_BGK}
 f_{i}^{\alpha} \left( \vec{x}+\vec{c}_{i} \Delta t, t+\Delta t \right) =
 f_{i}^{eq, \alpha} + \left( 1 -
 \frac{1}{ \tau_{mix} }  \right) f_{i}^{' \alpha} + \mathcal{F}_{i}^{\alpha}.
\end{equation}
The function $f_{i}^{'\alpha}$ is the nonequilibrium particle distribution for each fluid component. 
If $f_{i}^{'\alpha}$ takes the standard BGK form $f_{i}^{\alpha}-f_{i}^{eq, \alpha}$ and $\tau_{mix}$ is away from 1, one suffers from the instability caused by unphysical noise and numerical artifacts of the LB model.
To address this issue, a collision procedure regarding $f_{i}^{'\alpha}$ is regulated by,
\begin{equation}
\label{eq:Regualize}
f_{i}^{' \alpha}=\Phi^{\alpha}:\Pi^{\alpha}.
\end{equation}
Here $\Phi$ is a regularization operator that uses Hermite polynomials and $\Pi^{\alpha}$ is the nonequilibrium part of the momentum flux. The basic concept of regularized collision procedure can be found in \cite{2006_Chen, 2006_Zhang, 2006_Latt, 2006_Xiaowen_JFM,1997_Chen}. 

% 
% Boundary condition
%

To better achieve noslip wall boundary condition on arbitrary geometries, an extension of the volumetric boundary condition proposed by Chen et al in 1998 \cite{1998_Chen,2009_Leo,2004_Yanbing,2006_Fan} is utilized. In this method, after boundary surfaces are discretized into linear surface facets in two dimension or triangular polygons in three dimension, the incoming and outgoing particles based on those facets or polygons are calculated in a volumetric way obeying the conservation laws. 
The extension of this method for various types of the boundary conditon on arbitrary geometry has been studied. More details can be found in \cite{1998_Chen}. 
To reduce numerical smearing in near surface region, especially when physical viscosity is small and resolution is coarse, surface scattering model presented in \cite{2009_Leo} is crucial.
Our simulations of the single component flow have demonstrated the superior of this boundary condition model at low viscosity and resolution. Results will be reported in a separate paper.

To realize surface wetting conditions, the inter-component interaction force in Eq.~(\ref{eq:comp_interaction}) is extended to the interaction force between wall and fluid particles, $\vec{F}_{w}^{\alpha,  \beta}$, as,
\begin{eqnarray} 
\label{eq:wall_interaction}
\vec{F}_{w}^{\alpha, \beta} \left( \vec{x} \right) = G \rho_{\alpha} \left( \vec{x} \right) \sum_{i} w_{i} \vec{c}_{i} \rho^{'}_{\beta} \left(  \vec{x}+ \vec{c}_{i} \Delta t \right) ,  
\end{eqnarray}
for $\alpha \ne \beta$ and $\vec{F}_{w}^{\alpha, \beta} \left( \vec{x} \right) =0$ for $\alpha = \beta$.
$\rho^{'}_{\beta}$ is equivalent to the fluid density $\rho_{\beta}$ in the bulk region. And near the wall it is computed using fluid density and assigned solid density $\rho_{\beta}^s$, which is the wall potential for controlling the contact angle. This computation is performed so that the transition of $\rho^{'}_{\beta}$ is smooth \cite{1998_Chen} and the local momentum is conserved in bulk regions.

This volumetric wettability scheme has sufficient isotropy for keeping a droplet stable on the inclined wall to fluid lattices \cite{2015_Otomo}. 
The wall potential for components, $\rho_{1}^s$ and $\rho_{2}^s$, are controlled with a single parameter $\rho^s$.
%defined as 
%\begin{eqnarray} 
%\label{wall_potential_def}
%\rho_{1}^s=-\rho_0 \rho^s \Theta (-\rho^s), \; \; \: \: \: \rho_{2}^s=\rho_0 \rho^s \Theta (\rho^s),
%\end{eqnarray}
%using a single parameter $\rho^s$ where $\Theta$ is the Heaviside function and $\rho_0$ is 1.0.

%%%%%%%%%%%%%%%%%%%%%%%%%%%%%%%%%%%%%%%%%%
%%%%%%%%%%%%%%%%%%%%%%%%%%%%%%%%%%%%%%%%%%
%%        Validation
%%%%%%%%%%%%%%%%%%%%%%%%%%%%%%%%%%%%%%%%%%
%%%%%%%%%%%%%%%%%%%%%%%%%%%%%%%%%%%%%%%%%%
%
%%%%
\section{Validations}
\label{sec:validation}
%%%%

In this section, using the LB models described in the previous section, a set of test cases, which are necessary for the engineering application, is conducted in various systems and conditions.
Numerical results are compared to previous studies and analytical solutions. 

\subsection{A static droplet in free space}

Simulation of a two-dimensional static droplet in free space is performed using various resolution, viscosity, and the interaction strength in order to check the surface tension, interface thickness, droplet roundness, spurious current and their viscosity dependences.

\begin{figure*}[htbp]
  \begin{center}
    \begin{tabular}{c}
      % 1
      \begin{minipage}{0.5\hsize}
        %\begin{center}
          \includegraphics[clip, width=7.5 cm]{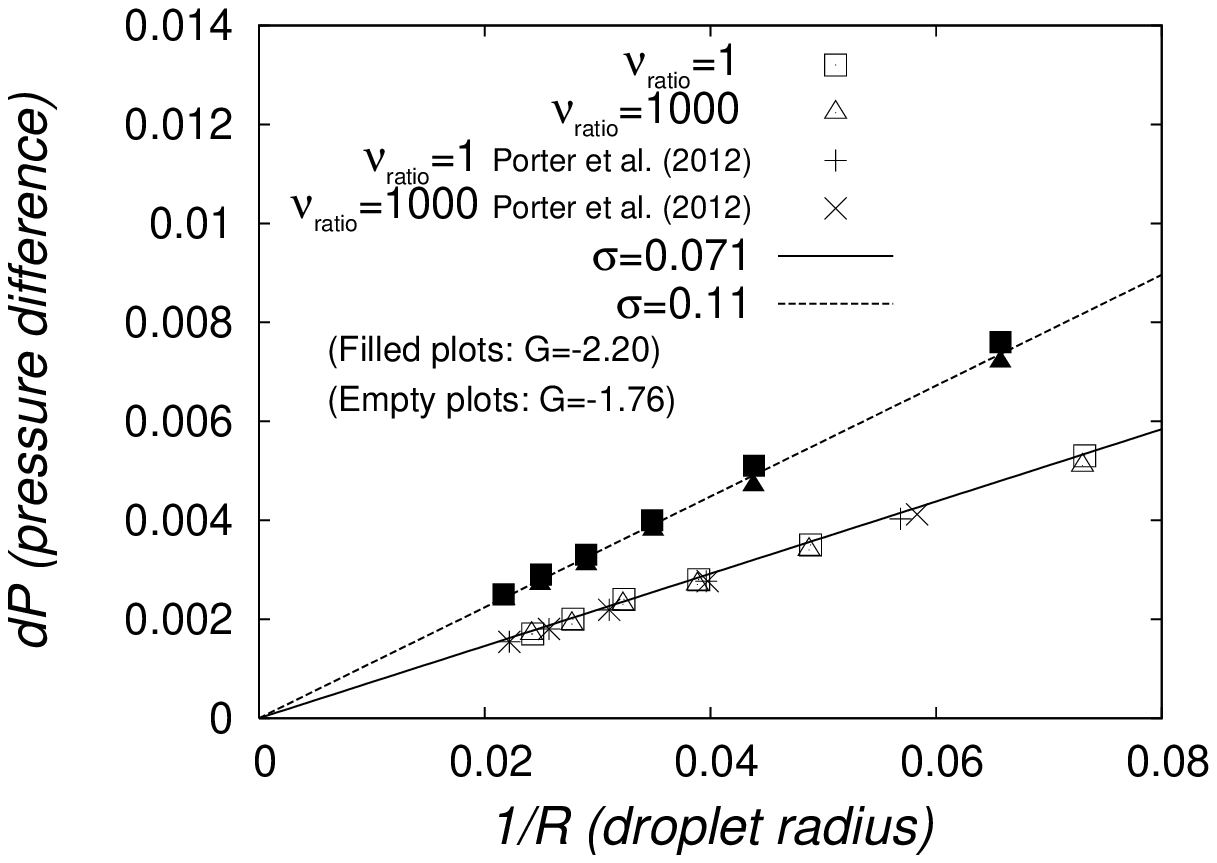}
        %\end{center}
	\caption{Pressure difference between the droplet (component 2) and suspended fluid (component 1) as a function of droplet radius with various viscosity, $G=-1.76$ (empty plots), and $G=-2.20$ (filled plots). Solid and dotted lines are ideal lines of $\sigma=0.073$ and $0.11$.}
	\label{fig:droplet_dp_vs_1R}
      \end{minipage}
      % 2
      \begin{minipage}{0.5\hsize}
        \begin{center}
          \includegraphics[clip, width=5 cm]{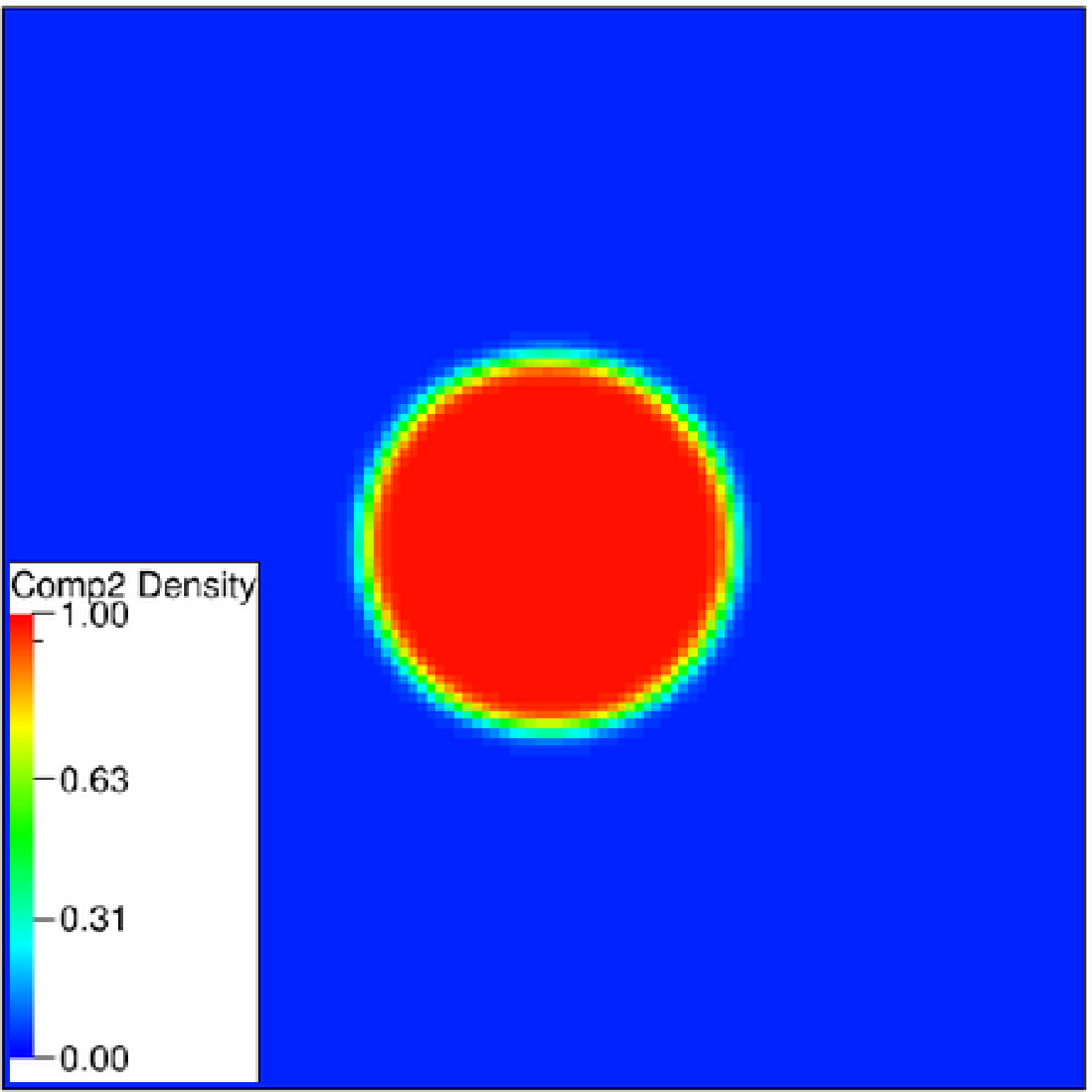}
        \end{center}
	\caption{Density contour of component 2 where $\nu_1=0.0017$, $\nu_2=1.7$, and $R=48$.}
	\label{fig:droplet_density_image}
      \end{minipage}
    \end{tabular}
  \end{center}
\end{figure*}

To measure the surface tension $\sigma$, a droplet with variable radius $R= \left\{ 16, \: 24, \: 30, \: 36, \: 42, \: 48 \right\}$ is initially set in the center of domain surrounded by periodic boundaries with edge lengths 5 times the droplet radius. Initial density for both components is $1.0$. The kinematic viscosity $\nu_1$ of component 1 and $\nu_2$ of component 2 are varied from $0.0017$ to $1.7$ independently, and the viscosity ratio $\nu_{ratio}= \nu_2 / \nu_1$ is varied up to 1000 as a result. Variable interaction strength defined in Eqs.~(\ref{eq:comp_interaction}) is $G= \left\{-1.72, -2.20 \right\}$. 
In Fig.~\ref{fig:droplet_dp_vs_1R} pressure difference between the droplet of the second component and suspended fluid of the first component is shown with respect to five viscosity choices. 
Plots are clearly distinguishable for each $G$ and results lie on solid and dotted lines which follow the Laplace law with $\sigma= \left\{ 0.071, 0.11 \right\}$ even with coarse resolution such as $R=16$. Furthermore the surface tension is almost independent on the viscosity whereas previous studies \cite{2011_Dong,2016_Otomo} reported that the original Shan-Chen model has significant viscosity dependence. 
Results in a study \cite{2012_Porter} are also presented in Fig.~\ref{fig:droplet_dp_vs_1R} and achieves similar improvements using the tenth order isotropy and multi-relaxation time scheme (MRT) at the sacrifice of computational cost and simplicity of implementation.

In Fig.~\ref{fig:droplet_density_image}, density contour of the second component where $\nu_1=0.0017$, $\nu_2=1.7$, and $R=48$ shows reasonable interface thickness and roundness of a droplet under extreme viscosity condition of $\nu_{ratio}=1000$.

%+++++ Figure (dP vs 1/R & droplet shape) ++++++
%+++++ Figure end +++++

\begin{figure*}[htbp]
  \begin{center}
    \begin{tabular}{c}
      % 1
      \begin{minipage}{0.5\hsize}
        \begin{center}
          \includegraphics[clip, width=8 cm]{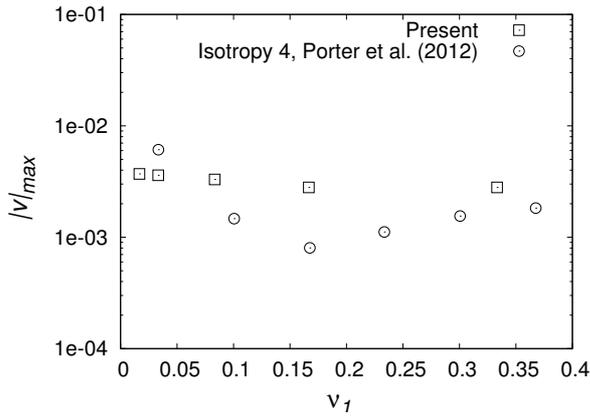}
        \end{center}
      \end{minipage}
      % 2
      \begin{minipage}{0.5\hsize}
        \begin{center}
          \includegraphics[clip, width=8 cm]{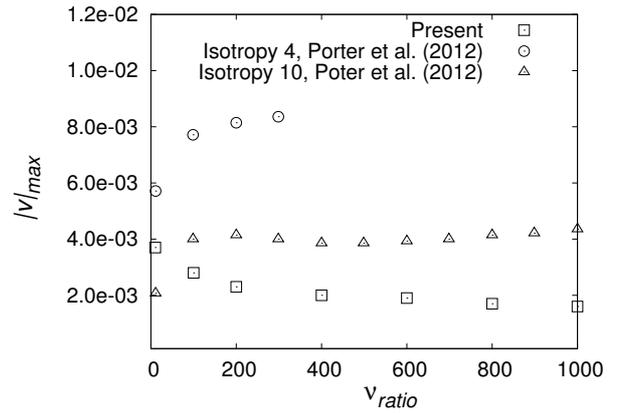}
        \end{center}
      \end{minipage}
    \end{tabular}
    \caption{Spatial maximum spurious current with respect to viscosity at $\nu_{ratio} = 1 $ (left) and to $\nu_{ratio}$ at constant suspended fluid's viscosity, $\nu_1=0.0067$. Results of LB models in \cite{2012_Porter} with the fourth and tenth order isotropy are plotted.}
    \label{fig:spu_current_vratio}
  \end{center}
\end{figure*}

The spurious current is investigated in terms of viscosity using a static droplet of $R=24$. In Fig.~\ref{fig:spu_current_vratio}, spatial maximum spurious current is plotted as a function of viscosity at $\nu_{ratio}= 1$ in the left figure and a function of $\nu_{ratio}$ at constant suspended fluid's viscosity $\nu_1=0.0067$ in the right figure. Results are compared to a previous study \cite{2012_Porter} in which the Shan-Chen model is improved using the tenth order isotropy, explicit forcing model, and the MRT scheme. Where $\nu_{ratio}=1$, viscosity dependence on the spurious current is less than their models and spurious current at low viscosity is slightly improved. Since for this case they used the same isotropy order as ours, the difference is possibly from the force model, relaxation time scheme, and regulated collision operator. 
Where $\nu_{ratio} \ne 1$, with their model and the fourth order isotropy, the simulation can be performed stably only with $\nu_{ratio}$ up to 300. 
On the other hand, with the same isotropy level, our model allows to simulate stably with $\nu_{ratio}$ up to 1000 and moreover has much reduced spurious current. Even compared to their model with the tenth order isotropy, the spurious current is improved at high $\nu_{ratio}$.
It is worth mentioning that the fourth order isotropy is beneficial in terms of computational costs since it requires substantially less stencils than the tenth order isotropy. Furthermore the treatment of boundaries is more straightforward with less stencils.

%+++++ Figure (Spurious current various visc ratio and visc) ++++++
%+++++ Figure end +++++

\subsection{Multi-component Poiseuille flow}

Multi-component layered flow driven by the gravity is simulated in a two-dimensional channel. For the channel height $64$/$128$, the gravity is $1.6e-7$ / $4.0e-8$, respectively. The first component is mainly occupied along the wall and the second component is in the middle of the channel.

\begin{figure*}[htbp]
  \begin{center}
    \begin{tabular}{c}
      % 1
      \begin{minipage}{0.5\hsize}
        \begin{center}
          \includegraphics[clip, width=8 cm]{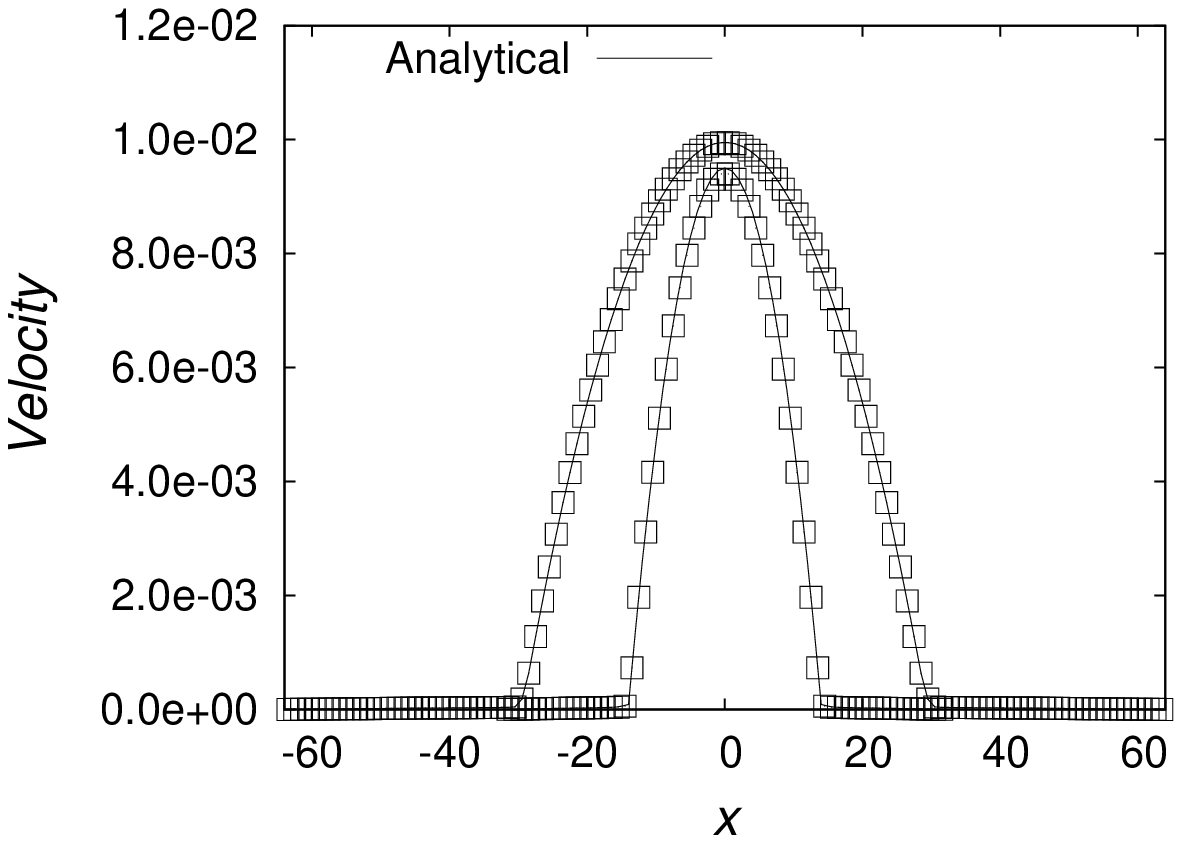}
        \end{center}
      \end{minipage}
      % 2
      \begin{minipage}{0.5\hsize}
        \begin{center}
          \includegraphics[clip, width=8 cm]{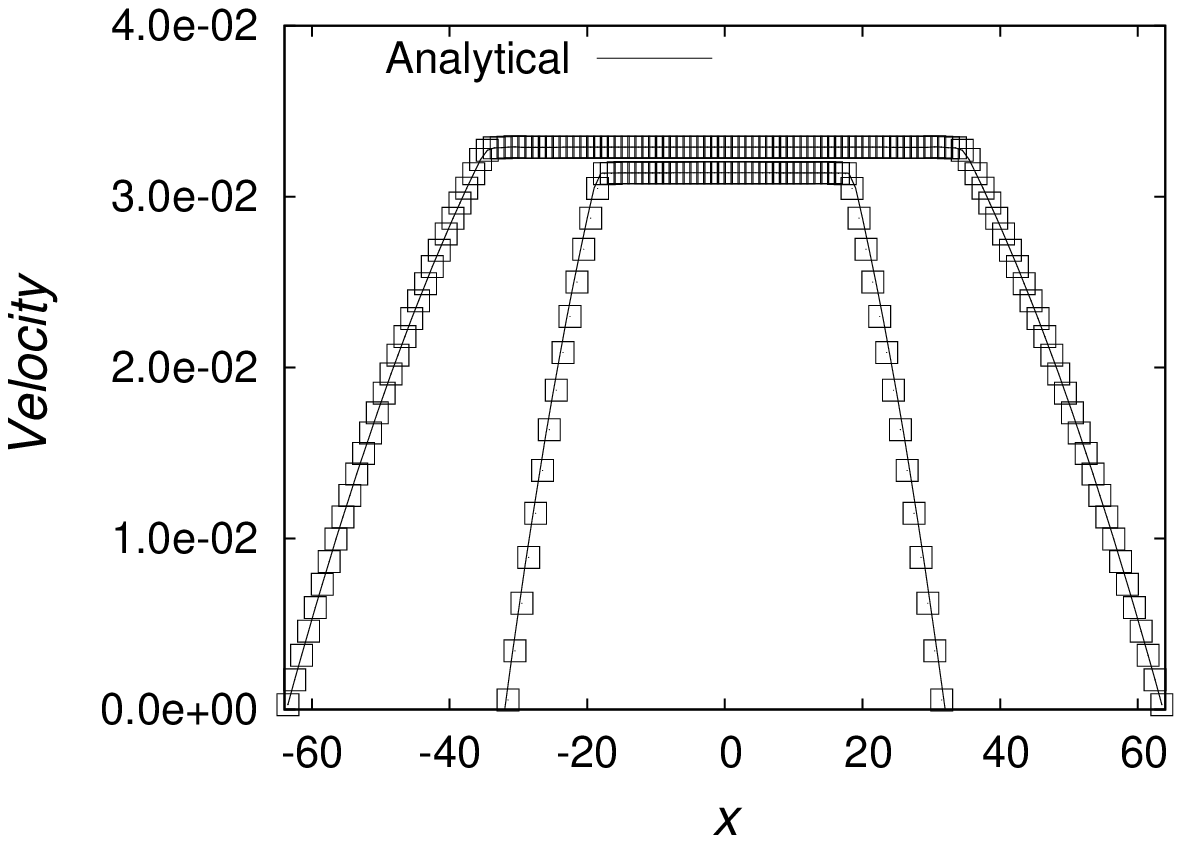}
        \end{center}
      \end{minipage}
    \end{tabular}
    \caption{Velocity profile at $\left(\nu_1, \nu_2 \right)=\left( 1.7,  0.0017 \right) $ in the left figure and $\left( 0.0017, 1.7 \right) $ in the right figure. The first component is along the wall and the second component is in the middle of the channel. Results with channel heights 64 and 128 are plotted in each figure. Solid lines are analytical solutions.}

    \label{fig:multi_poseuille}
  \end{center}
\end{figure*}

In Fig.~\ref{fig:multi_poseuille}, numerical results are compared to analytical solutions where $\left( \nu_1, \nu_2 \right) =\left( 1.7,  0.0017 \right) $ and $\left( 0.0017, 1.7 \right) $. In analytical solutions, the density weighted viscosity shown in Eq.~(\ref{mix_nu}) is considered. All simulated results agree with analytical solutions very well. In a study \cite{2012_Porter}, for obtaining accurate velocity profile in these setups, the MRT model and the tenth order isotropy scheme are required. However the LB model in this study can yield accurate results with the single-relaxation time model and the fourth order isotropy. Moreover compared to the other multi-component LB models in the paper \cite{2013_Yang}, in which accurate velocity profile can be obtained with $\nu_{ratio}$ up to 5 for the Shan-Chen model and up to 120 for the free-energy model and the color-gradient model, the LB model in this study shows much better accuracy and stability.

%+++++ Figure (Co-current layered flow) ++++++
%+++++ Figure end +++++

\subsection{A static slug between flat plates}
\label{subsec_contact}

A static slug in a two-dimensional channel is simulated for measurement of the relation between the contact angle $\theta$ and the wall potential at various viscosity. A slug is mainly composed of the second component and filled in half of the domain which is bounded by periodic boundaries in the direction along the wall. The rest of space is occupied by the first component.
The channel height and length are set as 32 and 128, respectively. 
Details of the contact angle measurement method can be found in \cite{2015_Otomo}. 

\begin{figure*}[htbp]
  \begin{center}
    \begin{tabular}{c}
      % 1
      \begin{minipage}{0.5\hsize}
        \begin{center}
          \includegraphics[clip, width=8 cm]{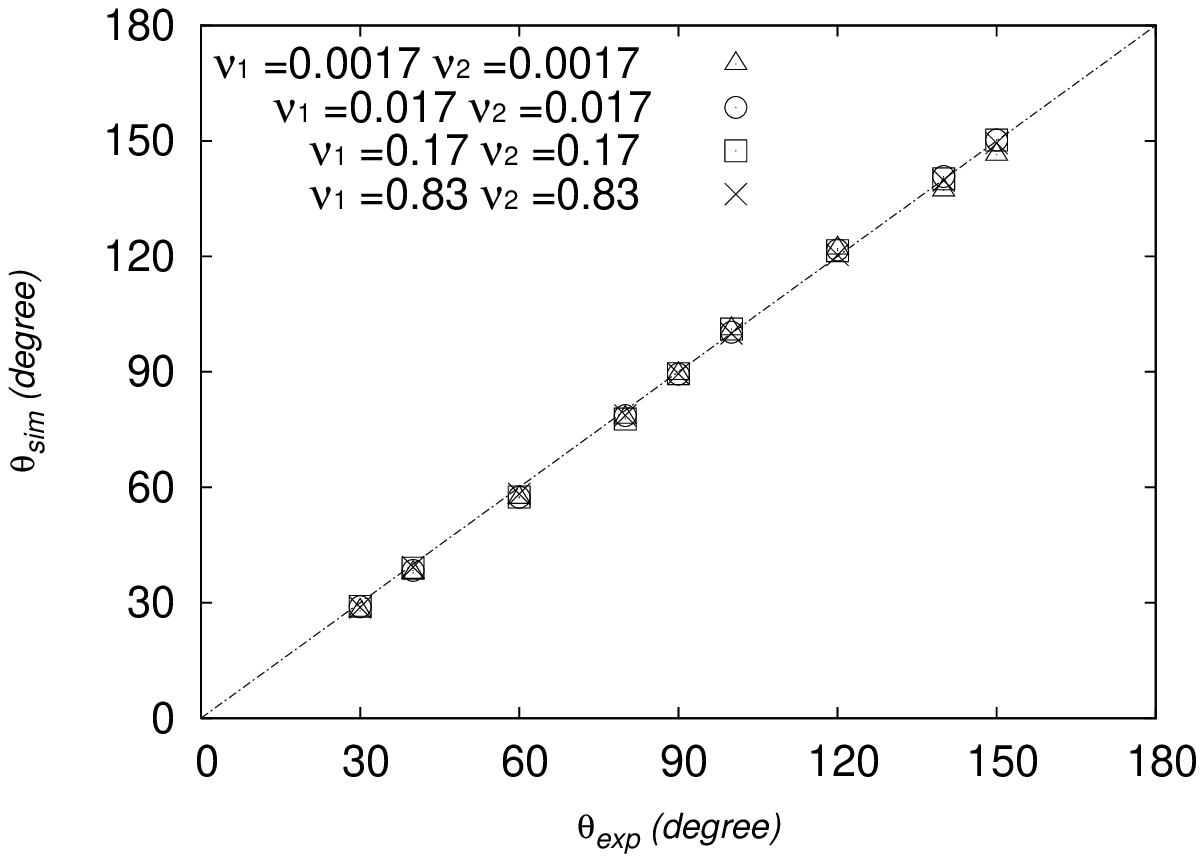}
        \end{center}
      \end{minipage}
      % 2
      \begin{minipage}{0.5\hsize}
        \begin{center}
          \includegraphics[clip, width=8 cm]{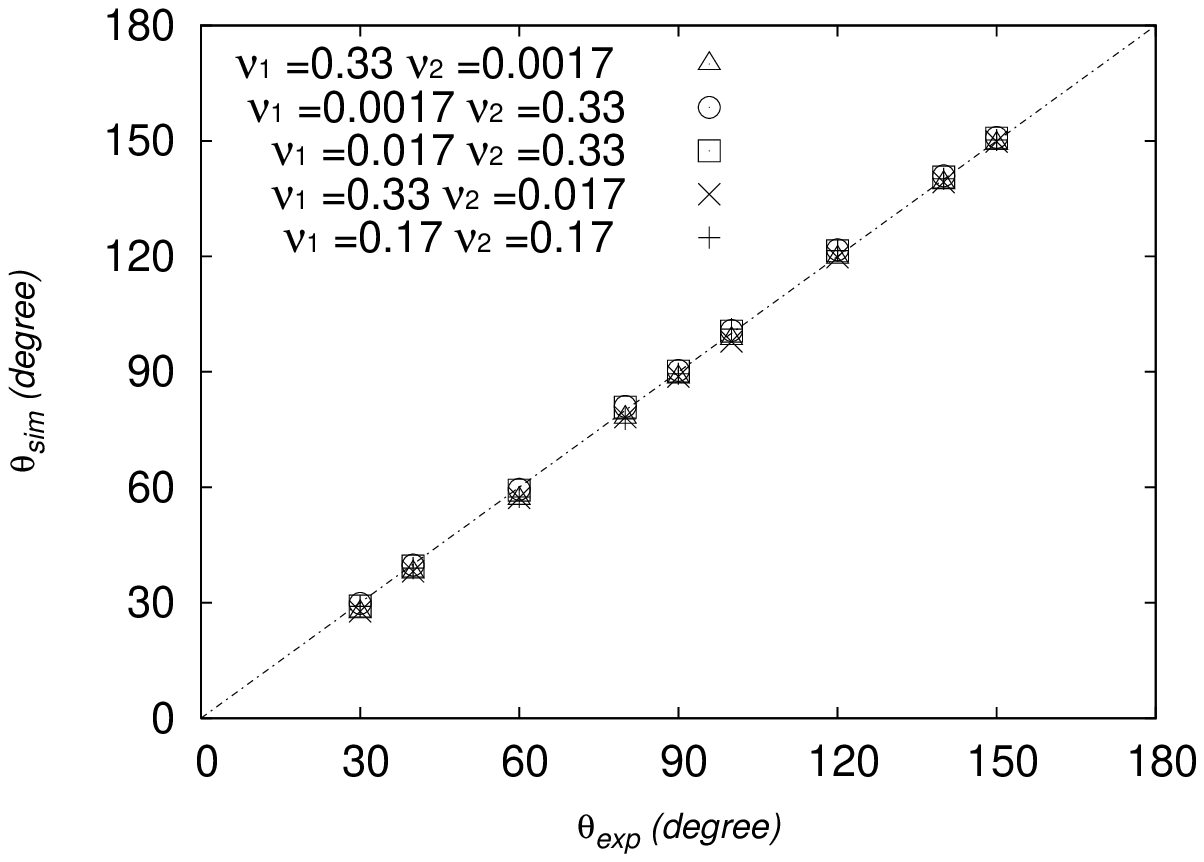}
        \end{center}
      \end{minipage}
    \end{tabular}
    \caption{ Simulated contact angle $\theta_{sim}$ as a function of the expected contact angle $\theta_{exp}$ using sets of $\nu_1$ and $\nu_2$ of $\nu_{ratio}=1$ in the left figure and $\nu_{ratio} \neq 1$ mainly in the right figure.}
    \label{fig:slug_cont}
  \end{center}
\end{figure*}

By improvements of the lattice condition dependence on $\theta$ in \cite{2015_Otomo}, accurate $\theta$ can be easily measured. Using numerical results, we formulate $\theta$ as a function of $\nu_1$, $\nu_2$, and $\rho^{s}$, namely $\theta= L \left( \nu_{1}, \nu_{2}, \rho^{s}  \right)$. To invert this function $L$ enables us to find a wall potential $\rho^{s}$ for a certain set of $\theta$, $\nu_{1}$, and $\nu_{2}$. For checking the accuracy of this procedure, using $\rho^s$ evaluated by sets of $\nu_1$, $\nu_2$, and the expected contact angle $\theta_{exp}$, a slug between plates is simulated. In Fig.~\ref{fig:slug_cont}, measured contact angles of a slug are shown for each set of variables.
The figure shows that for a variety of viscosity options, the contact angle can be simulated very accurately.

%+++++ Figure (Cotact angle slug) ++++++
%+++++ Figure end +++++

\begin{figure*}[htbp]
  \begin{center}
    \begin{tabular}{c}
      % 1
      \begin{minipage}{0.5\hsize}
        \begin{center}
          \includegraphics[clip, width=8 cm]{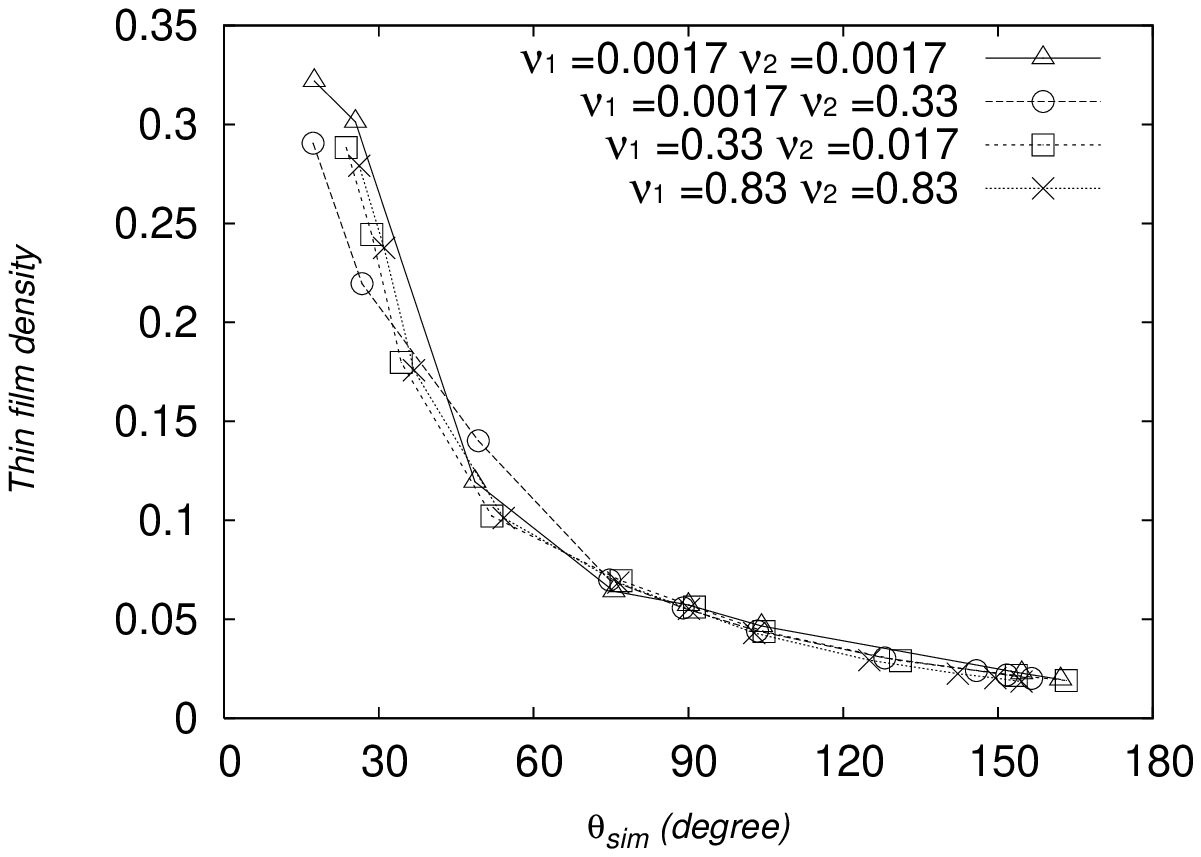}
        \end{center}
      \end{minipage}
      % 2
      \begin{minipage}{0.5\hsize}
        \begin{center}
          \includegraphics[clip, width=8 cm]{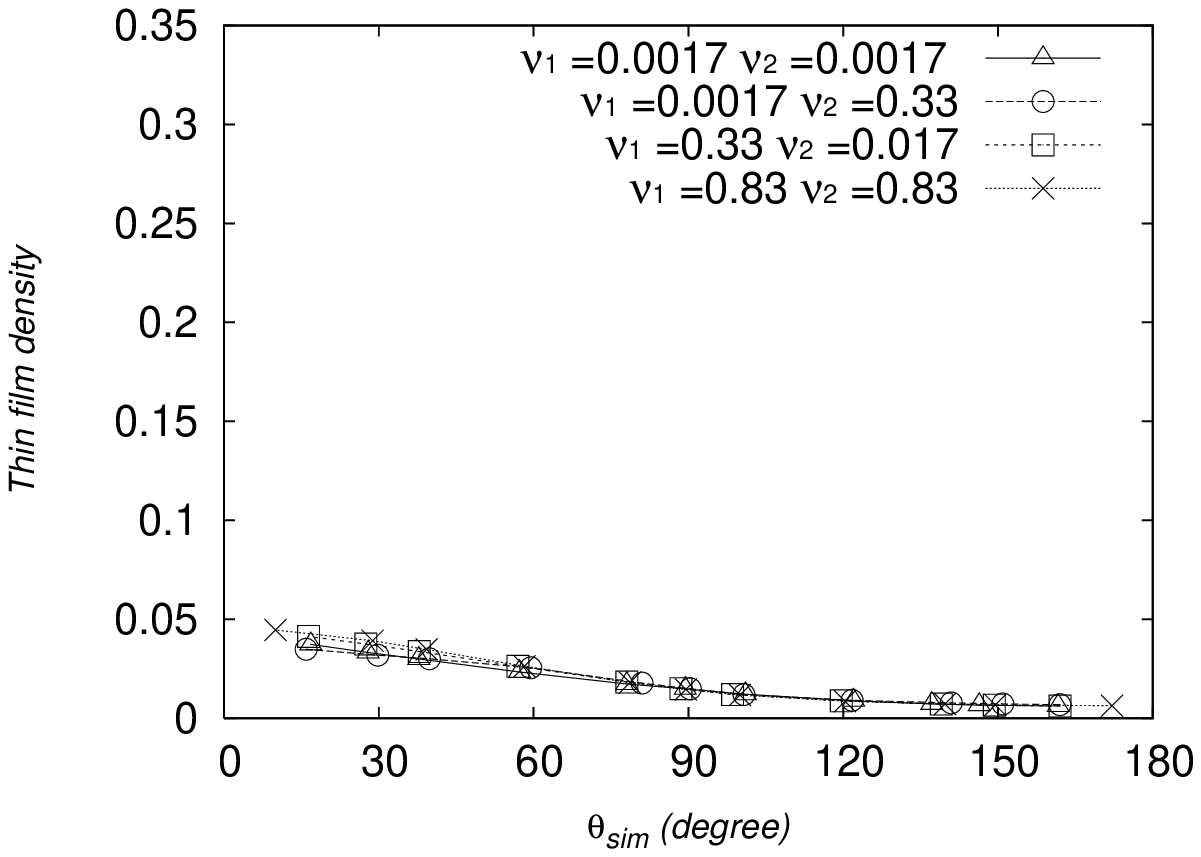}
        \end{center}
      \end{minipage}
    \end{tabular}
    \caption{ Thin film density as a function of the contact angle with original wetting model (left) and modified wetting model (right) using a variety of viscosity options.}
    \label{fig:thin_film}
  \end{center}
\end{figure*}

Where a wall is wetting for a certain fluid component, the fluid tends to be strongly diffusive and constitutes thin film, which is artificially thick for the macroscopic fluid dynamics, along the wall. This can be serious problems for engineering application such as mass leaking from the inlet and outlet, imprecise fluid distribution,  and inaccurate friction force along walls due to the intercomponent force.
As discussed in \cite{2015_Otomo}, this artificial film can be suppressed by the correction of surface wall potential. Here implementing the correction we tested with more various viscosity options.
In Fig.~\ref{fig:thin_film}, thin film density evaluated in the corner of domain are compared between original and improved models in terms of viscosity and contact angle. 
It shows that the thin film is reduced significantly at any viscosity and contact angles.

%+++++ Figure (Thin film) ++++++
%+++++ Figure end +++++

\subsection{A static droplet in inclined channels}

A static droplet in inclined channels to fluid lattices is simulated to study the lattice orientation dependency of the model. 
It has been a challenge to obtain a static droplet on a tilted surface without external force due to diffusion by the thin film along walls and insufficient isotropy of numerical algorithm near boundaries \cite{2016_Otomo}.
In this study the height of the channel is 16 and variable inclination angle is $\left\{ 30, 70 \right\}$  degree. 
A droplet is mainly composed of the second component for which the wall is wetting.
% A droplet is mainly composed of the second component which is hydrophilic. 

\begin{figure*}[htbp]
  \begin{center}
    \begin{tabular}{c}
      % 1
      \begin{minipage}{0.5\hsize}
        \begin{center}
          \includegraphics[clip, width=8 cm]{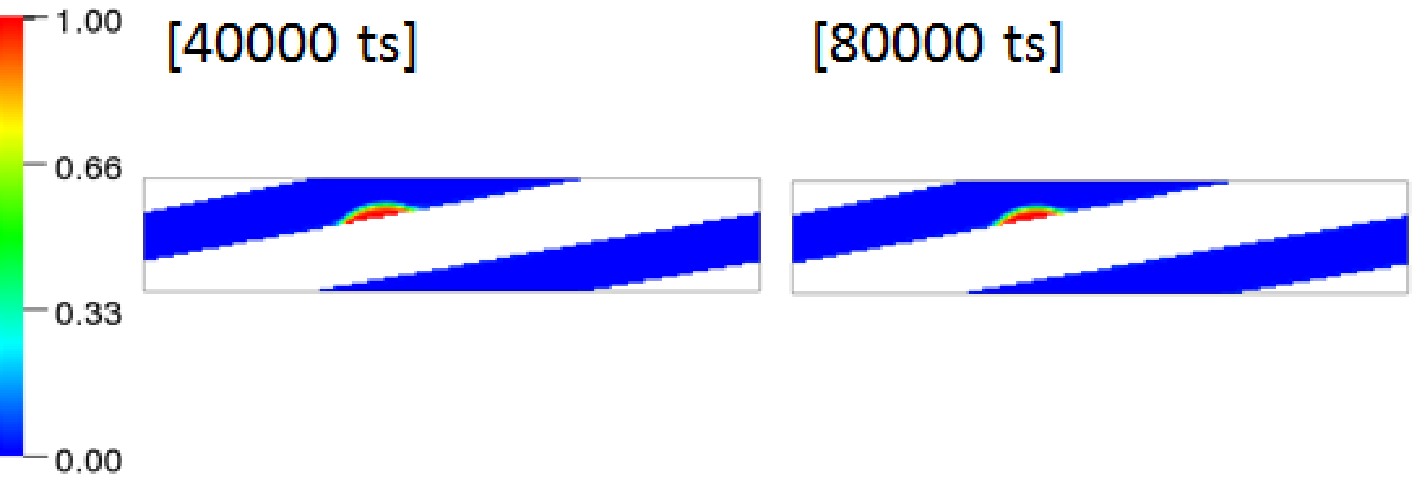}
        \end{center}
      \end{minipage}
      % 2
      \begin{minipage}{0.5\hsize}
        \begin{center}
          \includegraphics[clip, width=5 cm]{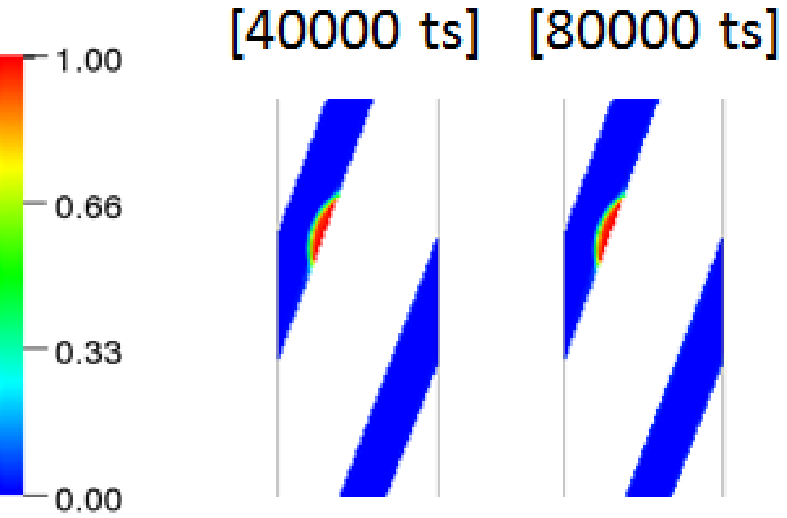}
        \end{center}
      \end{minipage}
    \end{tabular}
    \caption{ Density of the second component in the inclined channels whose inclination angles are 30(left) and 70(right) degree. Sets of viscosity are $\left( \nu_{1}, \nu_{2} \right) =\left( 0.0017, 0.33 \right)$ (left) and $\left( \nu_{1}, \nu_{2} \right) = \left( 0.33, 0.0017 \right)$ (right).}
    \label{fig:droplet_incl}
  \end{center}
\end{figure*}

As shown in our previous study \cite{2016_Otomo}, the diffusion and artificial movement of a droplet are improved by the same wetting model under many conditions of viscosity and wall inclination angles.
Here similar improvements are observed for more extensive viscosity options. Fig.~\ref{fig:droplet_incl} shows using $\left( \nu_{1}, \nu_{2} \right) = \left( 0.0017, 0.33 \right)$ and $\left( \nu_{1}, \nu_{2} \right) = \left( 0.33, 0.0017 \right)$ in which $\nu_{ratio}$ is 200 and 0.005, the droplet is static in a position without artificial diffusion.

%+++++ Figure (A static droplet on the inclined wall) ++++++
%+++++ Figure end +++++

\subsection{Permeability of $1 \%$ mixture flow}

Where a wall boundary is wetting for a certain component fluid, the fluid gets diffusive along walls with the film form as shown in Subsec. \ref{subsec_contact}. 
This film repels the other component fluid from near wall region due to the intercomponent interaction force and deteriorates the force imbalance between components, which leads to the artificial increase of the permeability.

%as a result reduces the friction on walls. For the macroscopic simulation, it leads to the artificial increase of the permeability.  

In this study, using the local momentum conserved interaction force discussed in Section.~\ref{sec:LB_formalism}, this issue is tried to be solved.  

\begin{figure*}[htbp]
\begin{center}
  \includegraphics[clip, width=12 cm]{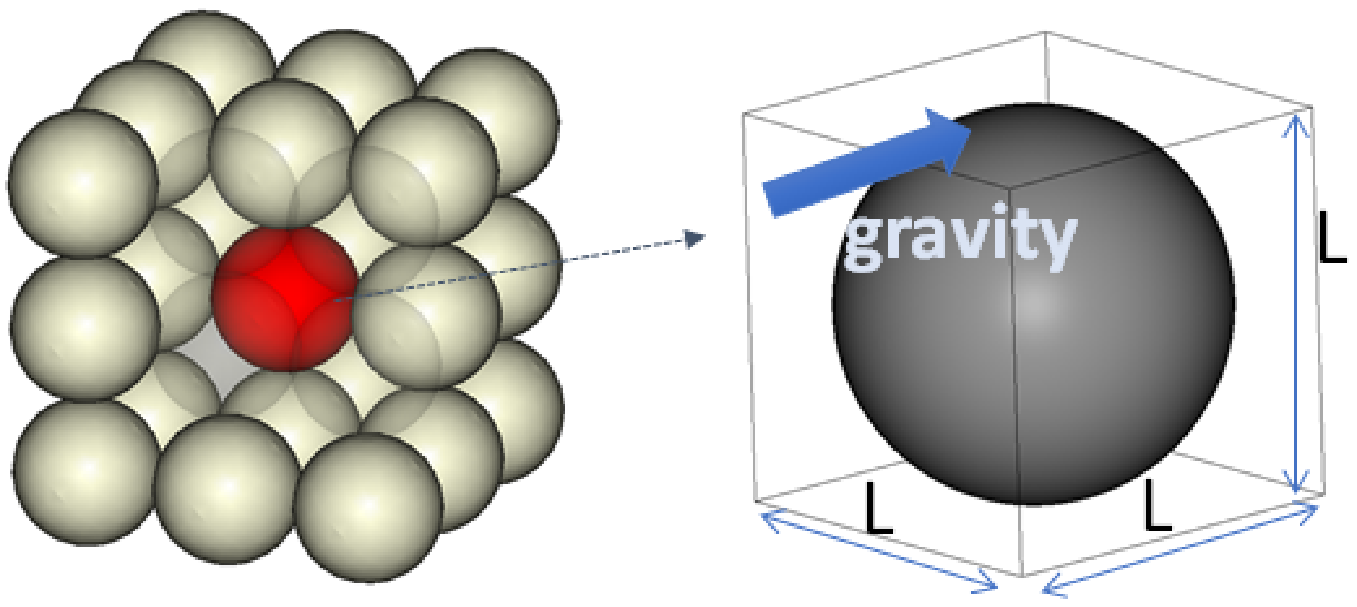}
\end{center}
\caption{Geometry of fully packed spheres array (left). In the computational model (right), domain is surrounded by periodic boundaries whose lengths are $L$. Over whole regions the gravity is applied.}
    \label{fig:geom_perm1permixed}
\end{figure*}

Using the fully sphere packed system shown in Fig. \ref{fig:geom_perm1permixed} which mimics porous media, two-component fluid flow is simulated. Here volume component ratio is 99 and viscosity ratio is 1. The sphere size $L$ in this figure is set as 20 and gravity $g$ is assigned so that $g \cdot \nu = 0.024$. 

%+++++ Figure (geometry) ++++++

%+++++ Figure end +++++

\begin{figure*}[htbp]
  \begin{center}
    \begin{tabular}{c}
      % 1
      \begin{minipage}{0.5\hsize}
        \begin{center}
          \includegraphics[clip, width=7 cm]{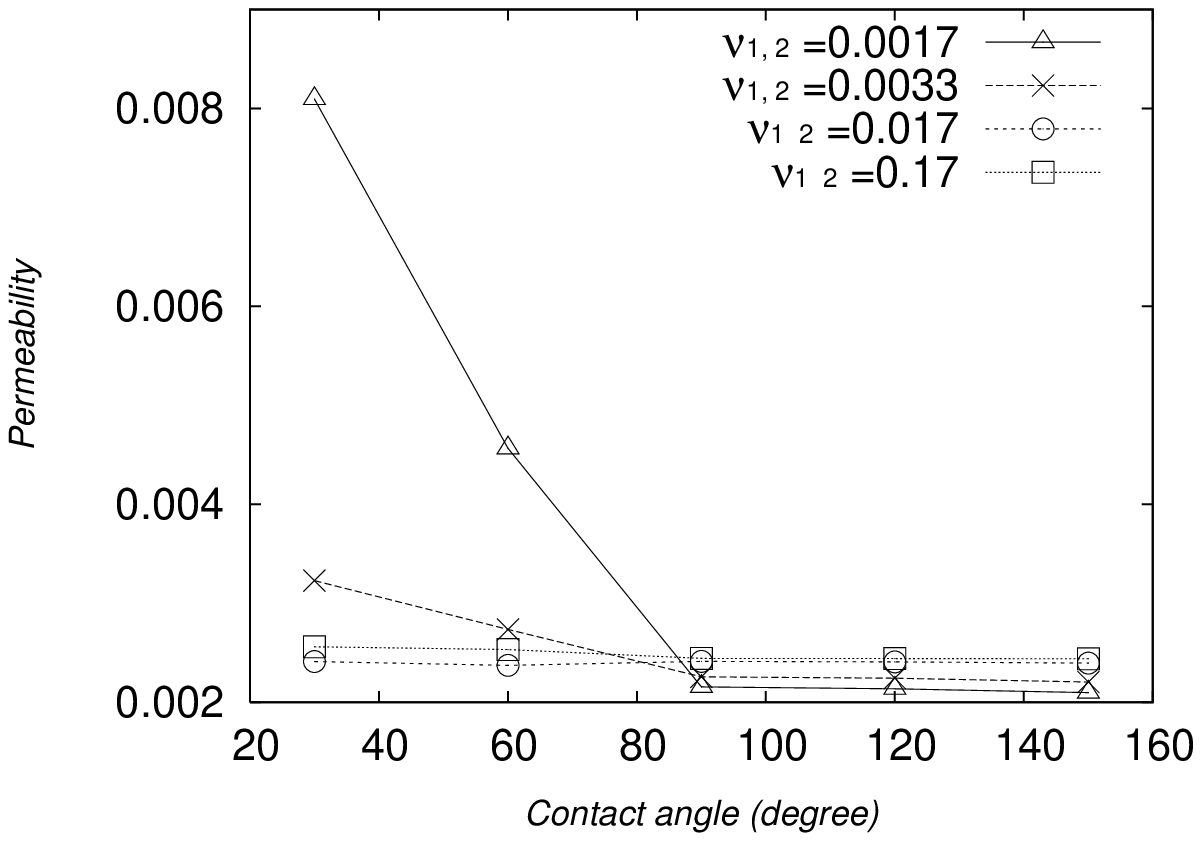}
        \end{center}
      \end{minipage}
      % 2
      \begin{minipage}{0.5\hsize}
        \begin{center}
          \includegraphics[clip, width=7 cm]{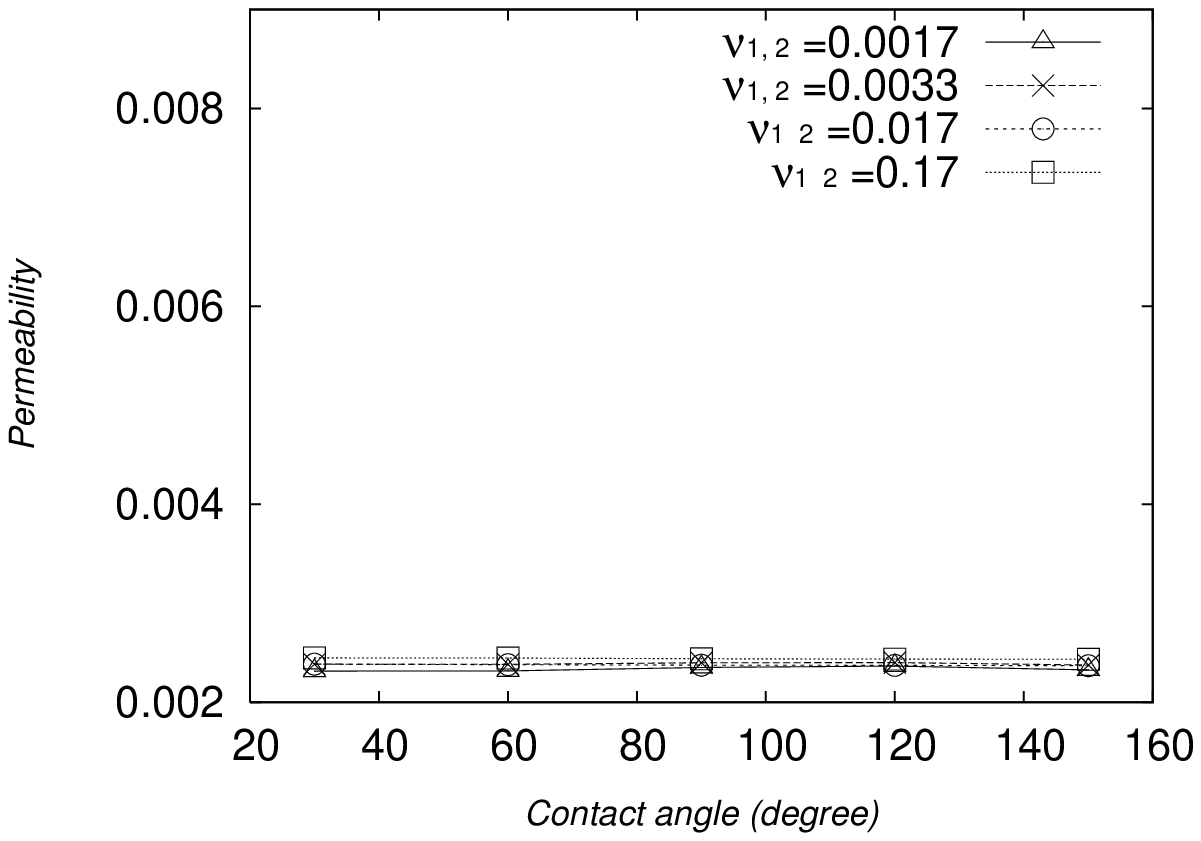}
        \end{center}
      \end{minipage}
    \end{tabular}
    \caption{Permeability of $1 \%$ mixture flow using various viscosity and contact angle using original wall model (left) and modified wall model(right). }
    \label{fig:perm1permixed}
  \end{center}
\end{figure*}

In Fig.~\ref{fig:perm1permixed}, permeability evaluated by $\nu U_D /g$, where $U_D$ is the Darcy velocity is shown using original wall model and current wall model with respect to various viscosity and contact angles. Since the flow is almost incompressible, using the porosity $\phi$ and the spatial averaged mixture velocity $U$ in the pore space, $U_D$ is approximated by $\phi U$.
The left figure of Fig.~\ref{fig:perm1permixed} shows that the original model results in significant increased permeability for stronger wetting and lower viscosity.
 This is because strong weattability makes the film on walls thicker and the lower viscosity makes velocity profile sensitive to the force imbalances due to the film. 
On the other hand, the right figure of Fig.~\ref{fig:perm1permixed} shows this issue is improved using the current model since forces along walls are exactly balanced between components.

%+++++ Figure (Permeability of 1% mixed flow) ++++++
%+++++ Figure end +++++

\subsection{Displacement of a slug in two-dimensional channels}

\begin{figure*}[htbp]
  \begin{center}
    \includegraphics[clip, width=12 cm]{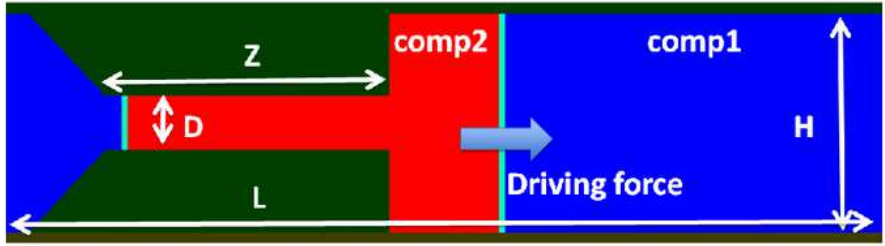}
    \caption{Geometry for testing the critical pressure}
    \label{fig:critical_prs_geom}
  \end{center}
\end{figure*}

In order to test the balance between capillary force and driving force, critical pressure for displacement of a slug from channels is measured.
In the previous study \cite{2015_Otomo}, similar tests are carried out using a sinusoidal channel regarded as a simple model of the porous media. Although numerical results were consistent with analytical solutions,
there was some uncertainty in analytical solutions because the slug volume cannot be evaluated accurately due to its nonzero interface thickness. To avoid this issue, a more simplified two-dimensional geometry shown in Fig.~\ref{fig:critical_prs_geom} is utilized in this study. 
%+++++ Figure (Geom: Displacement of a slug in 2D channel) ++++++
%+++++ Figure end +++++
Periodic boundaries are assigned on the left and right edges.
Channels with different heights $D$ and $H$ are connected and a slug composed of the second component is set over this connection. 
When $D<H$ and the wall is wetting for the second component, namely $\theta \le 90$ degree, there is a non-zero analytic critical pressure $P^{A}_{crit}$,
% When $D<H$ and the second component is hydrophilic, namely $\theta \le 90$ degree, there is a non-zero analytic critical pressure $P^{A}_{crit}$,
%
\begin{eqnarray}
\label{cirt_prs_form}
P^{A}_{crit}=2 \sigma cos \theta \left( \frac{1}{D} - \frac{1}{H} \right),
\end{eqnarray}
where $\sigma$ is the surface tension. Note that $P^{A}_{crit}$ doesn't depend on positions of interfaces. 
Compared to the sinusoidal channel case, the current setups bring benefits in terms of (i) much reduced computational cost due to two-dimensional geometry and (ii) accurate comparison with analytical solutions. It enables us to examine with a variety of settings such as resolution, viscosity and the contact angle more easily.
Furthermore the contact angle, which is not straightforward to be measured with coarse resolution due to non-zero interface thickness and thin film along walls, can be evaluated for any resolution indirectly by measured critical pressure and Eqs.~(\ref{cirt_prs_form}). 
%Then to check whether the interface is spherical or not by the relation between capillary length and channel height is required.
%after a check of spherical interface by the relation between capillary length and channel height is carefully checked. 
%
It is necessary to check whether the interface is spherical or not by the relation between capillary length and channel height and it is indeed verified in current simulations.

The middle and total channel length shown in Fig.~\ref{fig:critical_prs_geom} are $Z=52$ and $L=160$. The rear channel height is fixed as $H=40$ and variable middle channel heights are $D =\left\{5, 10 \right\}$.
The choices of the contact angle and viscosity are $\theta= \left\{ 20, 40 , 60, 80 \right\}$ degree and $ \left( \nu_1, \nu_2 \right)$ $= \left\{ \left( 0.0017, 0.33 \right), \left( 0.33, 0.0017 \right)  \right\}$.
The initial density for both components, $\rho_0$, is set to 1. To measure the critical pressure, driving force $\rho_0 g$ is assigned over the whole domain and ramped up in every 80k timesteps. In particular, the acceleration $g$ deviated from $g^{A}_{crit}$ by $\left\{-30,  -10, -5, 0, +5, +10, +30 \right\} \%$ are applied in each stage where $g^{A}_{crit}=P^{A}_{crit}/\left( \rho_0 L \right)$ to allow uncertainty of the contact angle by $\pm 2.5$ degrees.

\begin{figure*}[htbp]
  \begin{center}
    \includegraphics[clip, width=12 cm]{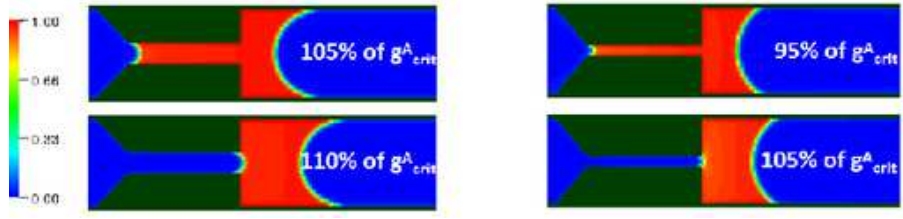}
    \caption{Density of the second component at $\nu_1=0.0017$ and $\nu_2=0.33$. The channel height $D$ is 10(left)/5(right) and the contact angle is 20(right)/40(left) degree. Images are taken when driving force, which is deviated from analytical solutions by $+5 \%$ and $+10 \%$ in the left figure and by $-5 \%$ and $+5 \%$ in the right figure, is imposed.}
    \label{fig:crit_prs_one_exampl}
  \end{center}
\end{figure*}

In Fig.~\ref{fig:crit_prs_one_exampl}, results at $\nu_1=0.0017$ and $\nu_2=0.33$ are shown. In the left and right figure, $D$/$\theta$ is defined as 10/20 degree and 5/40 degree, respectively. Those images are taken at timesteps when $g$ is deviated from $g^{A}_{crit}$ by $+5 \%$ and $+10 \%$ in the left figure and by $-5 \%$ and $+5 \%$ in the right figure. Because the transition is slow, obvious hysteresis effects are not observed. By those results we conclude that at this setups the simulated critical pressure is deviated from analytical solutions at most by $10 \%$ for the left figure and $5 \%$ for the right figure.

In Table.~\ref{tab:crit_prs_sum}, results for the other setups are summarized. Deviation percentage from analytical solutions $g^{A}_{crit}$ are below $10 \%$ in all cases. Because higher $D$ leads to smaller $P^{A}_{crit}$ in Eqs.~(\ref{cirt_prs_form}), more detailed force balance is required and therefore it may not be necessary that higher $D$ yields the higher accuracy.
Consequently, results show that reasonable critical pressure can be measured even at extreme kinematic viscosity ratio of 200, at extreme low kinematic viscosity of 0.0017, and very coarse resolution of 5.

%+++++ Figure (Result: Displacement of a slug in 2D channel) ++++++
%+++++ Figure end +++++

%+++++ Table (Critical pressure result) ++++++++++++++
\begin{table*}[htbp]
  \begin{center}
    \caption{ Deviation percentage of simulated critical acceleration $g_{crit}$ from analytical solutions $g^{A}_{crit}$ for contact angles $\theta$ and channel heights $D$ at four viscosity combinations}
    \begin{tabular}{c}
      % 1
      \begin{minipage}{0.28\hsize}
        \begin{center}
%          \begin{table}[htb]
            \tabcolsep = 0.14cm
	\begin{tabular}{c c c} 
         \multicolumn{3}{c}{ $\begin{Bmatrix} \nu_1= 0.0017 \\ \nu_2=0.33  \end{Bmatrix}$} \\ \hline
	$\theta$    &  $D$    & Deviation    \\ 
	(degree)    &         & from $g^{A}_{crit}$ ($\%$)        \\  \hline
         20        &   5          &  0 - 5          \\ 
         40        &   5          &  0 - 5          \\ 
         60        &   5          &  0 - 5          \\ 
         80        &   5          &  0 - 5          \\ 
         20        &   10         &  5 - 10          \\ 
         40        &   10         &  0 - 5          \\ 
         60        &   10         &  0 - 5          \\          
         80        &   10         &  0 - 5          \\ \hline 
	\end{tabular}
%        \end{table}
        \end{center}
      \end{minipage}
      % 2
      \begin{minipage}{0.2\hsize}
        \begin{center}
%          \begin{table}[htb]
%            {
        \tabcolsep = 0.14cm
        \begin{tabular}{c} 
          $\begin{Bmatrix} \nu_1 = 0.33 \\ \nu_2 = 0.0017 \end{Bmatrix}$ \\ \hline
          Deviation \\ 
          from $g^{A}_{crit}$ ($\%$) \\ \hline
          0 - 5          \\ 
          5 - 10          \\ 
          0 - 5          \\
          0 - 5          \\
          0 - 5          \\
          0 - 5          \\
          5 - 10          \\ 
          0 - 5          \\  \hline
	\end{tabular}
        \end{center}
      \end{minipage}
      % 3
      \begin{minipage}{0.2\hsize}
        \begin{center}
%          \begin{table}[htb]
%            {
        \tabcolsep = 0.14cm
        \begin{tabular}{c} 
          $\begin{Bmatrix} \nu_1 = 0.0017 \\ \nu_2 = 0.0017 \end{Bmatrix}$ \\ \hline
          Deviation \\ 
          from $g^{A}_{crit}$ ($\%$) \\ \hline
          0 - 5          \\ 
          0 - 5          \\ 
          0 - 5          \\
          0 - 5          \\
          0 - 5          \\
          0 - 5          \\
          0 - 5          \\ 
          0 - 5          \\  \hline
	\end{tabular}
        \end{center}
      \end{minipage}
      % 4
      \begin{minipage}{0.2\hsize}
        \begin{center}
%          \begin{table}[htb]
%            {
        \tabcolsep = 0.14cm
        \begin{tabular}{c} 
          $\begin{Bmatrix} \nu_1 = 0.33 \\ \nu_2 = 0.33 \end{Bmatrix}$ \\ \hline
          Deviation  \\ 
          from $g^{A}_{crit}$ ($\%$) \\ \hline
          0 - 5          \\ 
          0 - 5          \\ 
          0 - 5          \\
          0 - 5          \\
          0 - 5          \\
          0 - 5          \\
          0 - 5          \\ 
          0 - 5          \\  \hline
	\end{tabular}
        \end{center}
      \end{minipage}
    \end{tabular}
    \label{tab:crit_prs_sum}
  \end{center}
\end{table*}
%+++++ Table end (A droplet vel and  droplet mass/total mass on the Inclined wall) ++++

%+++++ Figure (Geom: Displacement of a slug in 2D channel) ++++++
%\begin{figure*}[htbp]
%  \begin{center}
%    \begin{tabular}{c}
%      % 1
%      \begin{minipage}{0.5\hsize}
%        %\begin{center}
%          \includegraphics[clip, width=6 cm]{./Crit_prs_result.eps}theta_40_h_5_nut1_0005_nut2_10.eps}
%        %\end{center}
%      \end{minipage}
%      % 2
%      \begin{minipage}{0.5\hsize}
%        \begin{center}
%          \includegraphics[clip, width=6 cm]{./theta_20_h_10_nut1_0005_nut2_10.eps}
%        \end{center}	 
%      \end{minipage}
%    \end{tabular}
%  \end{center}
%  \caption{Density contour of component 2 where $\nu_1=0.00017$ and $\nu_2=0.33$ when two options of gravity are imposed. %The channel height $D$ is 5(left)/10(right) and the contact angle is 40(right)/20(left) degree.}
%  \label{fig:crit_prs_one_exampl}
%\end{figure*}
%+++++ Figure end +++++

\subsection{Viscous and capillary fingering}
According to experimental and numerical studies in porous media and simple channels \cite{1988_Lenormand,1994_Halpern,2002_Chin,2004_Kang,2010_Dong,2011_Dong,2011_Zhang,2013_Liu,2013_Yang}, it is well known that the fluid displacement pattern depends on the viscosity ratio and capillary number.
For an example, if the capillary number $Ca$ is sufficiently low and viscosity ratio $\nu_{ratio}$ , the ratio of displacing fluid's $\nu$ to displaced fluid's $\nu$, is high, then the dominant force is the capillary force and one has the displacement pattern called \emph{capillary fingering} in which intruding fluid preferably propagates to large throat regardless of driving force direction and tends to create large blobs. Whereas if $Ca$ is high and $\nu_{ratio}$ is low then the viscous force is dominant and yields the pattern called \emph{viscous fingering} which intrudes to the displaced fluid with stretched thin form. Conditions for those patterns were summarized with the phase diagram \cite{1988_Lenormand,2011_Zhang,2013_Liu}. 

\begin{figure*}[htbp]
  \begin{center}
    \includegraphics[clip, width=12 cm]{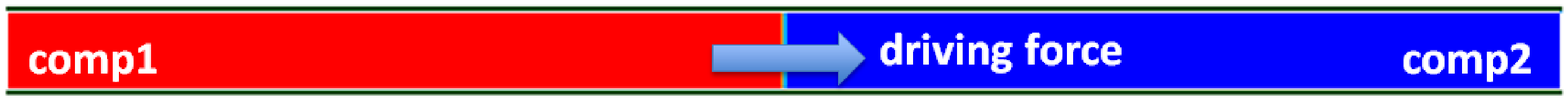}
    \caption{Setups for tests of the fingering}
    \label{fig:fingering_geom}
  \end{center}
\end{figure*}

In this study, those fingering patterns are simulated in a two-dimensional channel shown in Fig.~\ref{fig:fingering_geom}. The channel, whose height and length are 32 and 640, is bounded by periodic boundaries on its left and right edges. Two components are initially separated and after five thousand timesteps the driving force is imposed over the whole domain.

%+++++ Figure (Geom: Fingering) ++++++
%+++++ Figure end +++++

\begin{figure*}[htbp]
  \begin{center}
    \includegraphics[clip, width=15 cm]{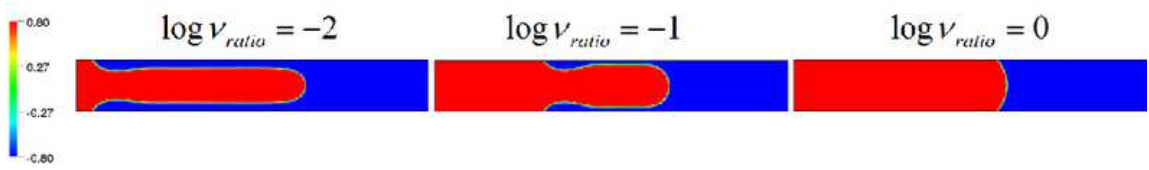}
    \caption{Viscous fingering with three viscosity ratios. The contour shows the Atwood number, $\left( \rho_1- \rho_2 \right)/\left( \rho_1+ \rho_2 \right)$. Those snapshots are taken when the tip velocity achieves 0.04.}
    \label{fig:viscous_fingering}
  \end{center}
\end{figure*}

Firstly, the viscous fingering is studied with variable $\nu_{ratio}=\nu_2 / \nu_1$ where $\nu_2$ is fixed as $0.017$. The wall is wetting for the second component, specifically the contact angle $\theta= 60$ degree. The initial density for both components is set as 1.
The driving force is assigned to obtain $Ca= \rho_2 \nu_2 U / \sigma=5.9e-3$ where $U$ denotes velocity of the interface. 
In Fig.~\ref{fig:viscous_fingering} displacement patterns are shown for three choices of $\nu_{ratio}$ with color contours of the Atwood number, $\left( \rho_1 - \rho_2 \right) / \left( \rho_1 + \rho_2 \right)$. As $\nu_{ratio}$ is increased, the fingering pattern is less pronounced, which is reasonable since effects of viscous force on the interface become weak. According to the phase diagram in \cite{2011_Zhang,2013_Liu}, for $Ca^{*}=Ca/cos \theta=1.2e-2$ which is the case in the current simulation, one expect that the viscous fingering starts to take place around ${\text log} \: \nu_{ratio}=-1$.  This expectation is consistent with Fig.~\ref{fig:viscous_fingering} although they estimated it in porous media and this transition is possibly influenced by details of geometry to some extent.
Compared to a study \cite{2004_Kang,2010_Dong} in which $\nu_{ratio}$ for stable simulation is limited only from 0.1 to 10, the LB model in this study allows us to use much more various $\nu_{ratio}$.

%+++++ Figure (Result: Viscous fingering) ++++++
%+++++ Figure end +++++

\begin{figure*}[htbp]
  \begin{center}
    \includegraphics[clip, width=15 cm]{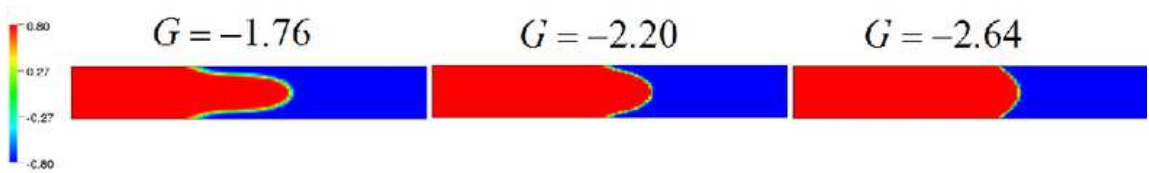}
    \caption{Capillary fingering with three inter-component interaction strengths $G$. The contour shows the Atwood number. The tips velocity in those figures is 0.024.}
    \label{fig:capillary_fingering}
  \end{center}
\end{figure*}

The surface tension $\sigma$ dependence on the fingering pattern is studied. Using three choices of the interaction strength, $G= \left\{ -1.76, -2.2, -2.64 \right\}$ defined in Eq.~(\ref{eq:comp_interaction}), $\sigma$ is varied as $\sigma= \left\{ 0.073, 0.11, 0.015 \right\}$, respectively.
The driving force is imposed so that the tip velocity achieves 0.024. The surface wetting condition is neutral, $\theta = 90$ degree, and the viscosity is $\left( \nu_1, \nu_2 \right)= \left( 0.25, 5.0 \right)$, namely $\nu_{ratio}=20$. As shown in Fig.~\ref{fig:capillary_fingering}, less surface tension yields more pronounced fingering pattern with smaller finger width. Although articles \cite{2002_Chin,2013_Yang} reported that the original Shan-Chen model failed to yield this behavior, the LB model in this study improves this issue. We believe that the enhanced force scheme and regulation of non-equilibrium terms mainly contributes to this improvement.

%+++++ Figure (Result: Capillary fingering) ++++++
%+++++ Figure end +++++

\begin{figure*}[htbp]
  \begin{center}
    \includegraphics[clip, width=15 cm]{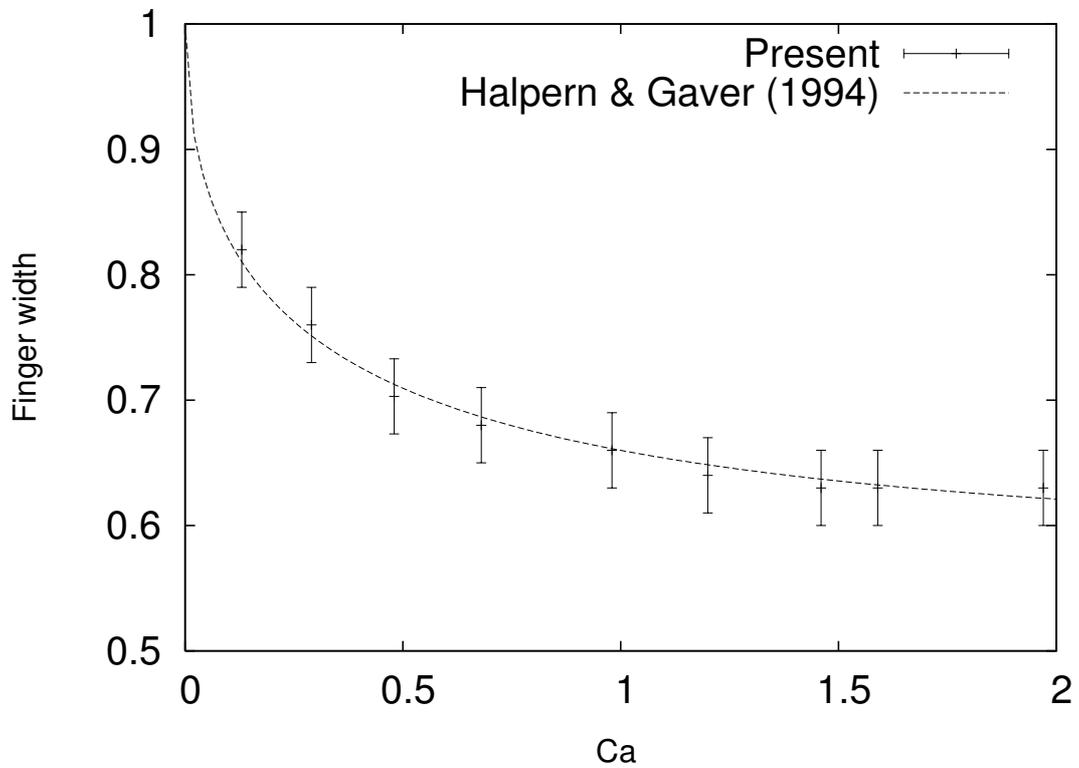}
    \caption{Finger width as a function of the capillary number. It is non-dimensionalized by the channel height. Error bars are originated from the interface thickness.}
    \label{fig:capillary_fingering_width}
  \end{center}
\end{figure*}

The finger width is evaluated with varied capillary numbers and compared to a previous study \cite{1994_Halpern}. 
Numerical finger width which is non-dimensionalized by the channel height is presented as a function of capillary number in Fig.~\ref{fig:capillary_fingering_width}. Error bars come from the interface thickness. Those plots agree with results of \cite{1994_Halpern} in which the analysis is based on the boundary element method. It is worth mentioning that the extended stable viscosity range enables us to simulate using such high capillary number. For large capillary number, the surface tension needs to be adjusted with attention to the interface thickness.

%+++++ Figure (Result: Capillary fingering width) ++++++
%+++++ Figure end +++++

%
%%%%%%%%%%%%%%%%%%%%
\section{SUMMARY}
\label{sec:summary}
%%%%%%%%%%%%%%%%%%%%

A multi-component lattice Boltzmann models for robust and accurate simulation at various viscosity is introduced and validated. Main features of our models are (i) a regulated collision operator which extracts the hydrodynamic mode from the non-equilibrium terms, (ii) the third order Maxwellian distribution in the equilibrium state, (iii) viscosity taking account of component mixture, and (iv) boundary models regarding friction and wettability based on a volumetric way.

Through tests in a variety of systems and conditions, the model is examined in the range of the kinematic viscosity $\nu$ from $0.0017$ to $1.7$ and kinematic viscosity ratio $\nu_{ratio}$ up to $1000$. Findings are itemized in follows,

\vspace{3mm}

\begin{itemize}

\setlength{\itemsep}{3mm} 
\item[$\bullet$] In simple systems such as a static droplet in free space and multi-component Poiseuille flow, in the range of $\nu$ from $0.0017$ to $1.7$ and $\nu_{ratio}$ up to $1000$, results are stable and accurate. In terms of spurious current, viscosity dependence on the surface tension, and accuracy of velocity profile in Poiseuille flow, outcomes are comparable or at some conditions superior to a previous study in \cite{2012_Porter} in which much more complicated LB models are applied. Velocity profiles in Poiseuille flow agree well with analytical solutions at $\nu_{ratio} \le 1000$ whereas in a study \cite{2013_Yang} stable results are obtained only at $\nu_{ratio} \le 120$ with the free-energy and the color-gradient model and $\nu_{ratio} \le 5$ with the original Shan-Chen model.

\setlength{\itemsep}{3mm} 
\item[$\bullet$] In the range of $\nu$ from $ 0.0017$ to $0.33$ and $\nu_{ratio}$ up to $200$, the contact angle is controlled well using the wall potential and 
the thin film is significantly reduced. Accuracy of permeability of $1 \%$ mixture flow is enhanced and a stable droplet on the inclined wall is observed. 
Moreover in displacement of a slug from channels with various channel height, wetting condition, and viscosity, the numerical critical pressure is consistent with analytical solutions within deviation less than $10 \%$. 

\setlength{\itemsep}{3mm} 
\item[$\bullet$] Viscous and capillary fingering in a two-dimensional channel are examined using a variety of viscosity and capillary numbers. At fixed capillary number, viscosity dependence on the fingering pattern is consistent with phase diagrams in \cite{2011_Zhang,2013_Liu} that were investigated in porous media experimentally and numerically. The reasonable surface tension dependence on the fingering pattern, which couldn't be obtained with previous version of Shan-Chen model \cite{2002_Chin}, is observed. Furthermore finger width at various capillary numbers agrees with numerical results in \cite{1994_Halpern} based on the boundary element method.

\end{itemize}

\vspace{3mm}

The simulation results and comparisons with analytical solutions and previous studies indicate that the LB model in this study is accurate and stable over a wide range of viscosity ratio conditions while retaining high computational efficiency and applicability to complex geometry.
In the future, authors wish to apply this model for more engineering subjects.

\end{document}